\documentclass[twocolumn,showpacs,preprintnumbers,amsmath,amssymb,superscriptaddress]{revtex4}
\usepackage{graphicx}% Include figure files
\usepackage{dcolumn}% Align table columns on decimal point
\usepackage{bm}% bold math
\usepackage{natbib}% natbib

\begin{document}

\title{Magnetic Field Control of Exchange and Noise Immunity in Double Quantum Dots}

\author{M. Stopa}
\email{stopa@cns.harvard.edu} \affiliation{Center for Nanoscale
Systems, Harvard University, Cambridge, MA 02138}

\author{C. M. Marcus}
\affiliation{Department of Physics, Harvard University, Cambridge, MA 02138}

\begin{abstract}
We employ density functional calculated eigenstates as a basis for
exact diagonalization studies of semiconductor double quantum dots,
with two electrons, through the transition from the symmetric bias
regime to the regime where both electrons occupy the same dot. We
calculate the singlet-triplet
splitting $J(\varepsilon)$ as a function of bias detuning $\varepsilon$ and
explain its functional shape with a simple, double anti-crossing
model. A voltage noise suppression ``sweet spot," where
$dJ(\varepsilon)/d\varepsilon=0$ with nonzero
$J(\varepsilon)$, is predicted and shown to be
tunable with a magnetic field $B$.
 \vspace{-0.75cm}
\end{abstract}

\maketitle

%introduction

The goals of computation and information processing at the quantum
level have stimulated the efforts of many researchers to
coherently manipulate a variety of elementary quantum systems. The
scope of these candidate systems is wide \cite{wide}. Advanced
fabrication technology and inherent scalability, however, make
semiconductor systems
especially promising. The
investigation of these systems has recently produced some auspicious
results \cite{Petta05,Hanson05,Johnson05,Hatano05,Elzerman04}. The
goal of these particular studies is to coherently manipulate and probe the spin
and charge state of a small (typically electron number $N$=1 or 2)
system using: (i) time-varying electric fields from pulsed gates;
(ii) charge sensors from nearby quantum point contacts (QPCs)
\cite{Elzerman03,Field93}; and (iii) externally applied magnetic
fields $B$. Attention has focused recently on the regime adjacent to
the degeneracy line between the double dot charge states ($N_L$,
$N_R$) = (1,1) and (0,2) \cite{Petta05}. Here $N_L$ and $N_R$ denote
the electron numbers on the left and right dots.

In Ref.~\cite{Petta05} the lateral gates confining the double dot were pulsed
to produce a controllable ``detuning" $\varepsilon$ of the potential
($\varepsilon$ is the potential
difference between left and right gates measured from the
degeneracy point of (1,1) and (0,2)), in order to first prepare two
electrons in a singlet state in one (say, the right) dot, separate
them into the two dots, and then recombine them in the right dot,
i.e., $(0,2) \rightarrow (1,1) \rightarrow (0,2)$. For the employed
gate voltages and dot level spacings, the recombination was
suppressed by Pauli blocking in the case where the (1,1) electron
is in a triplet \cite{Ono02}. The singlet-triplet
splitting, or exchange coupling $J(\varepsilon)$, the spin phase coherence time and
the damping of Rabi oscillations between singlet and triplet were
all thereby measured as functions of $\varepsilon$ at the separation
point, the inter-dot tunnel coupling $t$, and magnetic field $B$. As
$\varepsilon \rightarrow 0^-$, $J(\varepsilon)$ exhibited a rapid
rise such that $dJ/d\varepsilon$ increased with $J$. The
effect of large $dJ/d\varepsilon$ is to enhance sensitivity
to voltage noise. The damping of Rabi oscillations, whose frequency is
determined by $J(\varepsilon)$, appeared to increase with frequency, presumably due to
dephasing caused by this voltage noise.

\begin{figure}
\begin{center}
\includegraphics[scale=0.33]{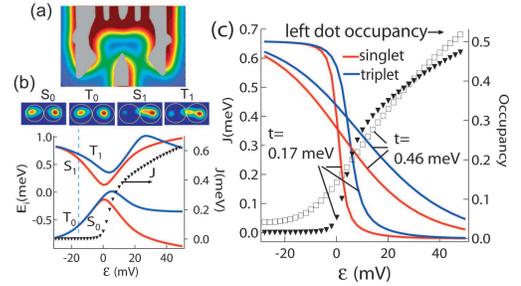}
\end{center}
\caption{(a) Self-consistent effective 2D potential profile at 2DEG
level, gate pattern superimposed; (b) Lowest two singlet and triplet
CI energies and difference of lowest triplet and singlet $J$
(triangles, right axis) versus detuning
$\varepsilon$, upper panel of (b) shows density of four states at
$\varepsilon = -18 \, mV$(dots
indicated) with $S_0$ and $T_0$ delocalized and $S_1$ and $T_1$
mostly in right dot; (c) $J(\varepsilon)$ for two (singlet) tunnel coupling
strengths $t$ (triangles and boxes) and lowest singlet and triplet
occupancies of left dot (right axis) for each $t$.} \label{Jdep}
\end{figure}

In this Letter we identify ranges of parameters where a noise-immunity
``sweet spot" in exchange can be found, that is, where exchange is present
($|J(\varepsilon)|>0$) but the system is insensitive to first order
electrical noise ($\partial J/ \partial \varepsilon = 0$). The
sweet spot is identified in configuration interaction (CI) calculations
\cite{Stopa05} for the electronic structure of the double quantum dot
with $N=2$ in Ref.~\cite{Petta05}. The CI calculation employs basis
states computed with density functional theory (DFT) \cite{Stopa96}
to obtain full geometric fidelity to the experimental structure.
We calculate $J(\varepsilon)$ from the symmetric limit at the center
of the (1,1) honeycomb in the stability diagram
well into the (0,2) regime with both electrons on a single dot. Analysis
within a Hartree-Fock (HF), double anti-crossing model allows us to
deduce simple expressions in the control parameters by which
the noise-immune regime can be accessed \cite{Hu06}.

DFT calculations for lateral heterostructures have been
described extensively in the literature \cite{Stopa96} . We
correct the DFT single particle energies of the N=2 double
dot to avoid double counting of the Coulomb interaction which is
diagonalized in the basis of Kohn-Sham states $\phi_i$
\cite{Stopa05}. The advantage of this method is that the basis
itself varies with gate voltages and $B$ and thereby captures much
of the evolving structure, much as a natural basis does in quantum
chemistry \cite{natural,Fulde}. The basis states are states of
the full double dot and so no artificial tunneling
coefficient needs to be incorporated \cite{tunneling}.
Finally, the Coulomb matrix elements automatically
include screening by the electrical environment and can be
calculated very efficiently with the kernel of Poisson's equation,
which is a natural byproduct of the DFT calculation.

The self-consistent potential profile from the DFT
calculation, with the device gate pattern superimposed, is shown in Fig.~1(a).
Here, the bias is approximately symmetric and the potential
minima of the two dots ($\sim 5 \, meV$ below the Fermi surface)
are nearly equal. Figure 1(b) displays the
CI-calculated lowest two singlet (S) and triplet (T) eigenstates as
a function of $\varepsilon$. The singlet and triplet ground states
each anti-cross with their corresponding first excited states (the
calculation preserves total spin). Also plotted is $J(\varepsilon)
\equiv E_T^- - E_S^-$, where $E_T^-$ and $E_S^-$ are the triplet and
singlet ground states, respectively. The nature of the four
anti-crossing states for $\varepsilon \approx -18 \, mV$ is shown by
the total (2D) density plots on the figure. Here, near the
anti-crossing, the two excited states on the (1,1) side each have
their densities concentrated in one dot, that is, they are the
states that become S and T (0,2) ground states for $\varepsilon>0$
\cite{anticross}. Fig. 1(c) exhibits $J(\varepsilon)$ for two
inter-dot tunnel-coupling strengths. Also shown are the S (red) and
T (blue) total occupancies of the \emph{left} dot versus
$\varepsilon$. The calculated results exhibit a coupling-dependent
rapid increase of $J(\varepsilon)$, as experimentally observed.
Additionally, in the (0,2) region ($\varepsilon>0$), where the $\varepsilon$
dependence of $J$ was not experimentally explored, the calculation
predicts a saturation.

\begin{figure}
\begin{center}
\includegraphics[scale=0.3]{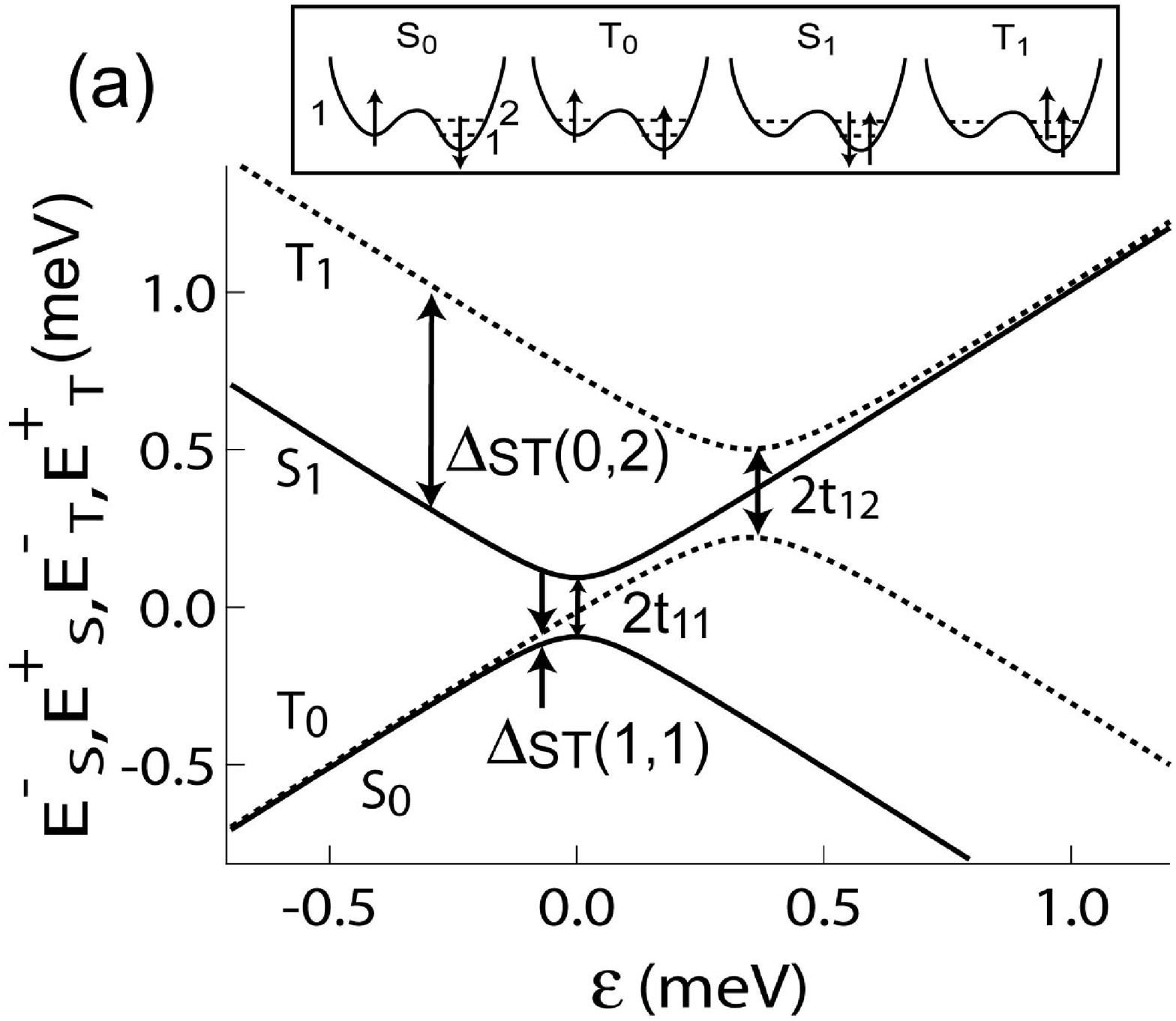}
\includegraphics[scale=0.3]{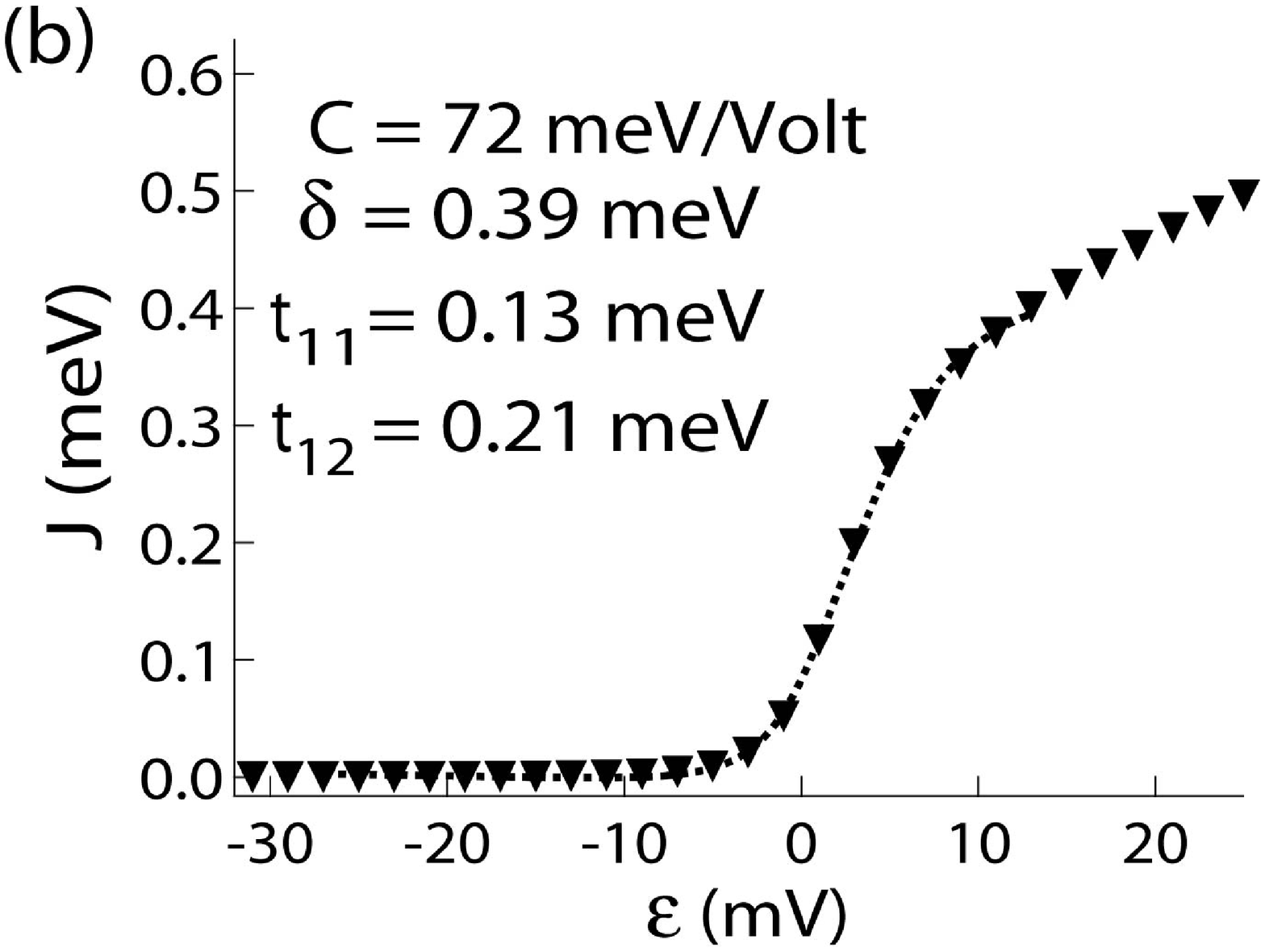}
\end{center}
\caption{(a) Schematic of double anti-crossing near (1,1) to (0,2)
transition. Singlets (solid) anti-cross with $t_{11}$, triplet (dashed) with
$t_{12}$. Singlet-triplet splitting of ground ($\Delta_{ST}(1,1)$)
and excited ($\Delta_{ST}(0,2)$) states in (1,1) region indicated. Top inset:
schematic of the configurations of the four anti-crossing states.
(b) Fit of calculated $J(\varepsilon)$ to anti-crossing equation
\ref{eq:J1} (dashed line) for $t=0.17 \, meV$ case in Fig. 1. Triangles
reproduce points in Fig. 1(c).}
\end{figure}

The structure of the numerical results in Fig.~1(b) suggest that we
can explain the shape of $J(\varepsilon)$ in terms of two overlapping
anti-crossings. These can be determined in their ideal form within a
simple Hartree-Fock (HF) approximation where the four ``bare" states are
written \cite{spectroscopic} as $S_0
\equiv (L1 \uparrow R1 \downarrow)$, $S_1 \equiv (R1 \uparrow R1
\downarrow)$ (singlets) and $T_0 \equiv (L1 \uparrow R1 \uparrow)$,
$T_1 \equiv (R1 \uparrow R2 \uparrow)$ (triplets). The single
particle energies are written $\epsilon_{R1}$, $\epsilon_{L1}$ and
$\epsilon_{R2}$ (we don't use $\epsilon_{L2}$). We include the
electrostatic potentials of the dot potential minima, $eC \phi_L$ and $eC
\phi_R$ in the bare level energies (here $C$ is a gate-to-dot lever
arm, assumed constant) hence $\epsilon_{L1} = \epsilon_{L1}^0 + e
C\phi_L$, where $\epsilon_{L1}^0$ denotes the energy measured from
the dot bottom. Similar expressions hold for the other level
energies. Within a HF type description, the bare
eigenfunctions are retained and the energies are shifted by
direct and exchange Coulomb matrix elements. Thus we write: $E_S^0 =
\epsilon_{L1} + \epsilon_{R1} + V_{inter}$, $E_S^1 = 2 \epsilon_{R1}
+ V_{intra}$, $E_T^0 = \epsilon_{L1} + \epsilon_{R1} + V_{inter} -
V_{inter}^{ex}$ and $E_T^1 = \epsilon_{R1} + \epsilon_{R2} +
V_{intra} - V_{intra}^{ex}$. For simplicity we here ignore the
state-dependence of the inter- and intra- dot matrix elements. We
also include the two (inter- and intra- dot) exchange
matrix elements only in the triplet terms. The anti-crossing of the
singlets results from the tunnel coupling of the two single particle
ground states which we
denote $t_{11}$ However, crucially, the triplets anti-cross with
tunnel coupling $t_{12}$, since the $R1$ state is blocked by the
Pauli principle. Due to the humped shape of the barrier, tunneling
from $L1$ to the higher $R2$ is stronger ($t_{12}>t_{11}$) and this
is evident in Fig.~1(b). The two branches of the singlet and triplet
are found by diagonalizing the standard 2x2 determinants:
\begin{equation}
\begin{split}
E_{S\pm}=& \tilde{E}_S \pm 0.5 \sqrt{(C\varepsilon)^2 + 4t_{11}} \\
E_{T\pm}=& \tilde{E}_T \pm 0.5 \sqrt{(\delta-C\varepsilon)^2
+4t_{12}} \label{eq:st}
\end{split}
\end{equation}
where $\tilde{E}_T \equiv (E_T^0 + E_T^1)/2$ and $\tilde{E}_S \equiv
(E_S^0+E_S^1)/2$. As noted above, $\varepsilon$ is by definition zero at the
singlet anti-crossing. In terms of the HF energies:
$C \varepsilon \equiv e C(\phi_L - \phi_R)
- \epsilon_{R1} - \epsilon_{L1} - V_{inter} + V_{intra}=0$.
Also, $\delta$ is the detuning (in units of energy, i.e. multiplied by
the lever arm $C$) at which the triplet anti-crosses, relative to
that of the singlet. It is given by $\delta
\equiv \epsilon_{R2}^0 - \epsilon_{R1}^0 - V_{inter}^{ex} +
V_{intra}^{ex}$. The exchange splitting is the difference between
the two lower branches in \ref{eq:st}:
\begin{equation}
J(\varepsilon) = \tilde{E}_T - \tilde{E}_S - 0.5 \sqrt{(C
\varepsilon - \delta)^2 + 4t_{12}^2} + 0.5\sqrt{(C\varepsilon)^2 +
4t_{11}^2} \label{eq:J1}
\end{equation}
note that $\tilde{E}_T - \tilde{E}_S$ is independent of
$\varepsilon$.

Figure 2(a) shows the energy for the few levels, $E_{S \pm}, E_{T \pm}$.
Away and to the left of the
anti-crossings, the gap between the ground states S and T is $E_T^0 -
E_S^0 \equiv \Delta_{ST}(1,1)=-V_{inter}^{ex}$ \cite{theorem}. The
gap between the (bare) excited states is $E_T^1 - E_S^1 \equiv
\Delta_{ST}(0,2)= \epsilon_{R2} - \epsilon_{R1} - V_{intra}^{ex}$,
i.e. it depends on a single particle level spacing in dot R. Clearly,
from Fig.~2(b), as $J$ increases near $\varepsilon=0$ so too does
$dJ/d\varepsilon$. For spin manipulation with noise immunity it would
be advantageous to find a regime where $J$ was appreciable but
$dJ/d\varepsilon$ was not. This turns out to be possible. In Fig.
2(b) we also fit $J(\varepsilon)$ from Fig. 1(c) (i.e. the full CI results,
specifically the curve with the smaller
dot-dot coupling) with Eq. \ref{eq:J1}. The fit is
good as long as points far into the (0,2) region are excluded
\cite{far02}.

In what follows
we show that the expression Eq.~\ref{eq:J1} has exactly one
extremum, $dJ/d\varepsilon = 0$, except where $t_{11}=t_{12}$. We
then show in what parameter region the extremum is a minimum and how the
parameters can be modulated by magnetic field $B$
to accentuate that minimum.

First, to show that $J(\varepsilon)$ has a single minimum, we
take the derivative of Eq.~\ref{eq:J1}:
\begin{equation}
\frac{dJ}{d\varepsilon} = -0.5 \frac{(C \varepsilon -
\delta)}{\sqrt{(C \varepsilon - \delta)^2 + 4 t_{12}^2}} + 0.5
\frac{C \varepsilon}{\sqrt{(C \varepsilon)^2+4 t_{11}^2}}
\end{equation}
and observe that the two terms are sigmoidal curves which (for
$t_{11} \ne t_{12}$) must intersect in one point, specifically
$\varepsilon_m = \delta/(1-(t_{12}/t_{11}))$. A second derivative
\begin{figure}
\begin{center}
\includegraphics[scale=0.3]{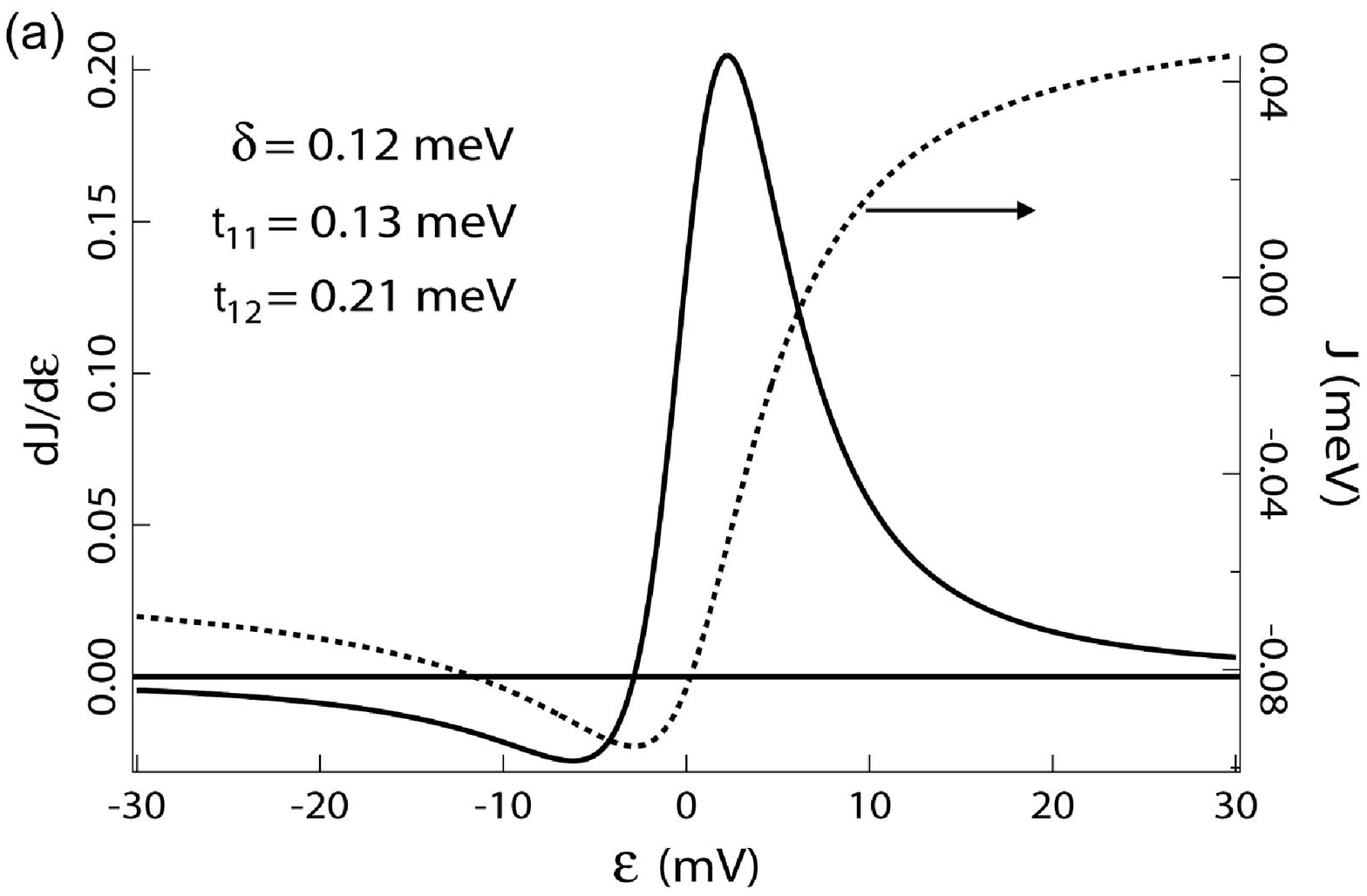}
\includegraphics[scale=0.3]{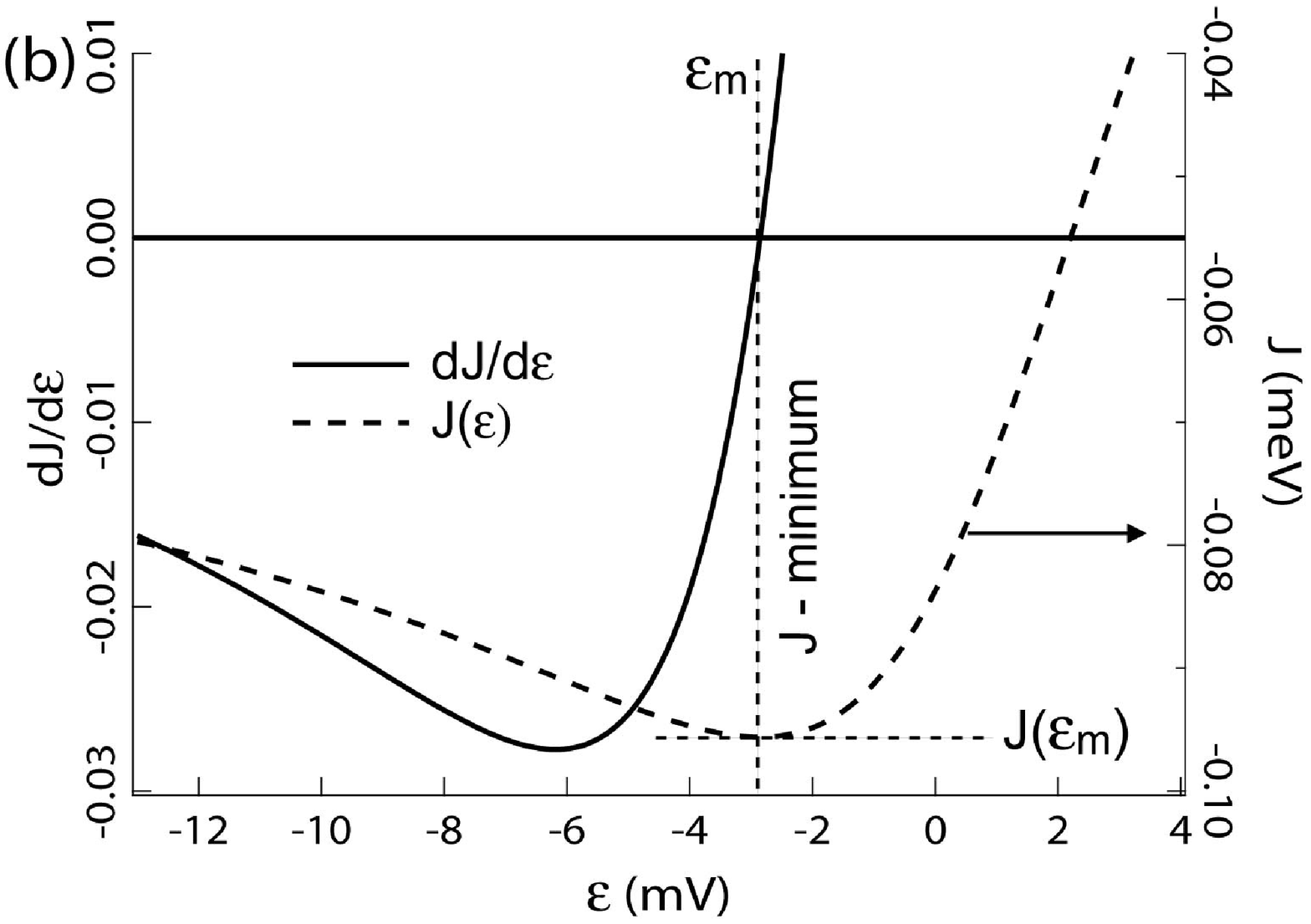}
\includegraphics[scale=0.3]{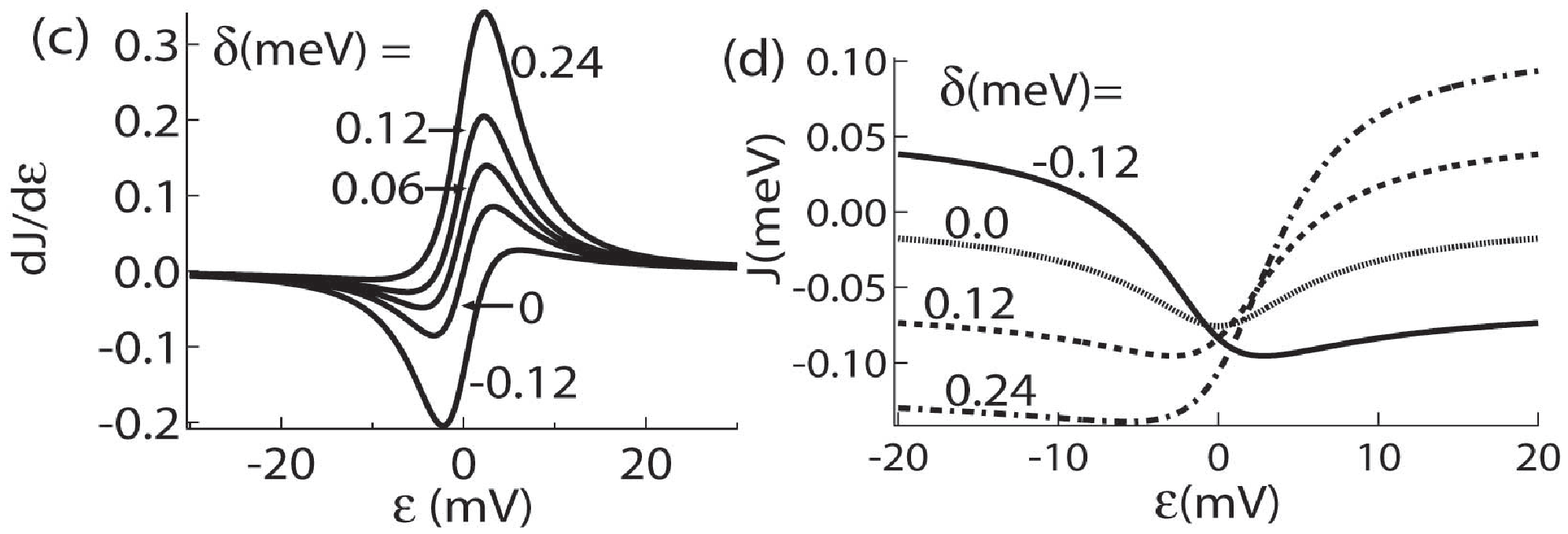}
\end{center}
\caption{$dJ/d\varepsilon$ (solid, left scale) and $J(\varepsilon)$,
from Eqs. 3 and 2, resp.,
for tunnel and $\delta$ parameters in Fig. 2(b). Note, offset energies,
$\tilde{E}_S$ $\tilde{E}_T$ taken as zero throughout figure, so offsets
of $J$ arbitrary. Expanded view in (b) shows
minimum ($dJ(\varepsilon_m)/d\varepsilon \equiv 0$) and depth of minimum. As
$\varepsilon \rightarrow - \infty$, $J$ saturates above the figure. For same
tunnel coefficients, $dJ/d\varepsilon$ (c) and $J(\varepsilon)$ (d), for
a range of values of $\delta$. Minimum of $J(\varepsilon)$
(i.e. $\varepsilon_m$) switches
sign to positive when $\delta$ changes sign to negative.}
\end{figure}
test shows that for $t_{12}>t_{11}$ (the usual case), the extremum
is a minimum. Note also that, assuming $t_{12}>t_{11}$, $\delta>0
\Rightarrow \varepsilon_m<0$ (the minimum is in the (1,1) zone) and
$\delta<0 \Rightarrow \varepsilon_m > 0$.

Evaluating the \emph{depth} $D$ of the
minimum of $J(\varepsilon)$ yields $D \equiv J(\varepsilon_m)- J(\varepsilon \rightarrow -
\infty) = 0.5 \delta(1-\sqrt{1 + 4 |t_{12}-t_{11}|^2/|\delta|^2}
\approx |t_{12}-t_{11}|^2/|\delta|$. The full CI calculations,
which include the realistic inter-dot barrier, allow us to estimate
the tunnel coefficients (compare Fig.~1(b) where the singlet and triplet
anti-crossings have gaps of $2t_{11}$ and $2t_{12}$, respectively).
Here, and in comparable parameter ranges, the tunnel
coefficient ratio $t_{12}/t_{11}\sim 1.3$.
From its definition, $\delta$ is the level spacing of dot R
corrected by inter- and intra-dot exchange energies and, at
B=0, is of order $1.2 \, meV$ from calculations for the
device in Ref. \cite{Petta05}.
Note that $\delta$ forms the major portion of $\Delta_{ST}(0,2)$
(cf. Fig. 2(a)) and that reducing $\Delta_{ST}(0,2)$ while keeping
$\Delta_{ST}(1,1)$ relatively fixed causes the two anti-crossings to
move closer to one another. For non-zero $B$ in a
circular parabolic potential, the lowest branch of the $2$ state
converges toward the $1$ state and the exchange term (Hund's
coupling) can induce a transition in the $N=2$ single dot to a
triplet ground state; Fig. 4(a) and Ref. \cite{Tarucha}. Simultaneously, $B$ induces a
singlet to triplet transition in the (1,1) ground state by enhancing
the wavefunction overlap in the saddle point
\cite{Stopa05,Burkard,Hu}, but the influence on the (0,2)
single particle levels is greater and the result is a drastic
decrease in $\delta$, which can even become negative.
\begin{figure}
\begin{center}
\includegraphics[scale=0.3]{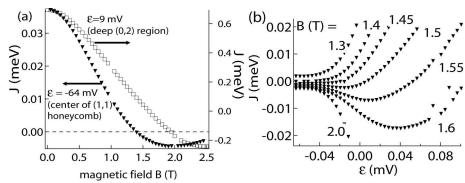}
\end{center}
\caption{Magnetic field dependence of $J$, from full CI calculations, for
center of (1,1) honeycomb cell (i.e. balanced gate voltages) triangles, and for
strong (0,2) regime where both electrons occupy right dot, boxes. Note
carefully difference between left and right scales. Single dot requires greater $B$ to drive S-T
transition, but $|dJ/dB|$ is much greater in (0,2). (b) Full CI calculation
for $J(\varepsilon)$ for various $B$ for strong coupling case in Fig. 1(c).
Effect of $B$ is principally to reduce single particle level spacing and thereby
reduce $\Delta_{ST}(0,2)$ (compare Fig. 2(a)). This moves
triplet anti-crossing to lower $\varepsilon$, i.e.
reduces $\delta$, simultaneously deepening the minimum of $J(\varepsilon)$.
When triplet anti-crosses before the singlet
($\delta<0$, $B=1.55 \, T$ and greater),
the minimum of $J(\varepsilon)$ occurs in the (0,2) zone ($\varepsilon>0$).}
\end{figure}

Figure 4(b) shows $J(\varepsilon)$ computed (full CI) for a variety
of $B$ values and for $t=0.46 \, meV$.
The minima increasingly deepen as a function of $B$ and move to larger $\varepsilon$.
Note that for this large value of tunnel coupling strength the
minimum for small $B$ (e.g. the $B=1.3 \, T$ curve)
is not observed at all. In this case the
tunnel coupling is comparable in scale to the level spacing and when
$\delta$ is large ($B$ is small) mixing of higher levels apparently
invalidates the simple, two-level HF picture which we have
used. The point is, however, that $t$ and $B$ can be varied to change
the order of singlet and triplet anti-crossings ($\delta$ goes from
$>0$ to $<0$). This shifts
the minimum of $J(\varepsilon)$ from (1,1) to (0,2) and, for $\delta=0$
the minimum can become very deep.

In summary,we have demonstrated that device parameters can be tuned
near the (1,1) to (0,2) crossover to generate a relatively noise-immune
sweet spot where the exchange $J(\varepsilon)$ exhibits a minimum.
The depth of the minimum,
$\sim |t_{12}-t_{11}|^2/|\delta|$,
and its detuning value, $\varepsilon_m = \delta/(1-(t_{12}/t_{11}))$,
can be modulated with magnetic field and tunnel
barrier height. The existence of
this noise-immune point for spin interactions is a promising development
for quantum dot quantum computation.

\indent We acknowledge the National Nanotechnology Infrastructure Network
Computation project, NNIN/C, for computational resources. We thank Edward Laird,
Amir Yacoby, Jason Petta and Jacob Taylor for helpful conversations.


\vspace{-0.5cm}
\begin{references}
\vspace{-1.5cm}

\bibitem{wide} Michael A. Nielsen and Isaac L. Chuang,
\emph{Quantum Computation and Quantum Information}, Cambridge
University Press, Cambridge, 2000; L. M. K. Vandersypen, Ph.D.
Thesis, Stanford University, July 2001.

\bibitem{Petta05} J. Petta {\it et al.}, \emph{Science} \textbf{309}, 2184
(2005).

\bibitem{Hanson05} R. Hanson \emph{et al.}, \emph{Phys. Rev. Lett.}
\textbf{94}, 196802 (2005).

\bibitem{Johnson05} A. C. Johnson \emph{et al.}, \emph{Nature}
\textbf{435}, 925 (2005); F. H. L. Koppens \emph{et al.}, \emph{Science}
\textbf{309}, 1346 (2005).

\bibitem{Hatano05} T. Hatano, M. Stopa and S. Tarucha, \emph{Science}
\textbf{309}, 268 (2005); T. Hatano, M. Stopa, T. Yamaguchi, T. Ota,
K. Yamada and S. Tarucha, \emph{Phys. Rev. Lett.}  93, 066806 (2004).

\bibitem{Elzerman04} J. M. Elzerman \emph{et al.} \emph{Nature}
\textbf{430}, 431 (2004).

\bibitem{Elzerman03} J. M. Elzerman \emph{et al.} \emph{Phys. Rev.
B} \textbf{67}, 161308 (2003).

\bibitem{Field93} M. Field \emph{et al.}, \emph{Phys. Rev. Lett.} \textbf{70},
1311 (1993).

\bibitem{Ono02} K. Ono, G. D. Austing and Y. Tokura, \emph{Science}
\textbf{297}, 1313 (2002).

\bibitem{Stopa05} W. van der Wiel, \emph{et al.}, New Journal of
Physics \textbf{8}, 28 (2006).

\bibitem{Stopa96} M. Stopa, Phys.Rev. B \textbf{54}, 13767 (1996); M.
Stopa, Semicond. Sci. Technol. {\bf 13}, A55 (1998).

\bibitem{Hu06} X. Hu and S. Das Sarma, Phys. Rev. Lett. \textbf{96}, 100501 (2006).

\bibitem{natural} P. L\"{o}wdin, Phys. Rev. \textbf{97}, 1474, 1955.

\bibitem{Fulde} P. Fulde,
{\it Electron Correlations in Molecules and Solids},
(Springer, Berlin, Heidelberg 1995).

\bibitem{tunneling} Our analytical treatment with a Hartree-Fock model does,
however, employ tunnel coefficients.

\bibitem{anticross} Note that at the left of Fig. 1(b), which is near
the center of the (1,1) stability cell, the excited S and T states
become degenerate. At this point the excited states are not (0,2)
states but are rather formed from orbitals above the first orbital in
each dot.

\bibitem{spectroscopic} We use ``$1$" and ``$2$" to denote the lowest
two single particle levels localized in one or the other dot.

\bibitem{theorem} The bare triplet can be lower than the singlet in
the approximation where tunnel coupling is ignored.

\bibitem{far02} Here the potential confinement begins to change and
the bare level spacing is no longer constant.

\bibitem{Tarucha} S. Sasaki et al., Nature (London) 405, 764 (2000).

\bibitem{Burkard} G. Burkard, D. Loss, and D. P. DiVincenzo, Phy. Rev. Lett. {\bf
59}, 2070(1999).

\bibitem{Hu}
X. Hu, and S. Das Sarma, Phy. Rev. A {\bf 61}, 062301(2000); M. Stopa,
Physica E {\bf 10}, 103 (2001).

\end{references}
\end{document}